\documentclass{article}
\usepackage{graphicx} 
\usepackage[title]{appendix}
\usepackage{authblk}

\usepackage{indentfirst,csquotes}
\usepackage[font={small}, labelfont=bf]{caption}
\usepackage{subcaption} 
\usepackage{multicol}
\usepackage{hyperref}

\title{\bf \Huge Cultural evolution in populations of Large Language Models}

\author[,1]{\bf  Jérémy Perez\thanks{Corresponding author: \texttt{jeremy.perez@inria.fr}}}
\author[1]{\bf Corentin Léger}
\author[2]{\bf Marcela Ovando-Tellez}
\author[2]{\bf Chris Foulon}
\author[3]{\bf Joan Dussauld}
\author[1]{\bf Pierre-Yves Oudeyer}
\author[1]{\bf Clément Moulin-Frier}

\affil[1]{Flowers Team, INRIA, Bordeaux, France}
\affil[2]{GIN-IMN, CNRS, Bordeaux, France}
\affil[3]{Independent Researcher}

\date{}

\begin{document}

\maketitle

\renewenvironment{abstract}{%
\begin{center}%
        {\bfseries \Large\abstractname}%
      \end{center}%
      \quotation}

\begin{abstract}

Research in cultural evolution aims at providing causal explanations for the change of culture over time. Over the past decades, this field has generated an important body of knowledge, using experimental, historical, and computational methods. While computational models have been very successful at generating testable hypotheses about the effects of several factors, such as population structure or transmission biases, some phenomena have so far been more complex to capture using agent-based and formal models. This is in particular the case for the effect of the transformations of social information induced by evolved cognitive mechanisms. We here propose that  leveraging the capacity of Large Language Models (LLMs) to mimic human behavior may be fruitful to address this gap. On top of being an useful approximation of human cultural dynamics, multi-agents models featuring generative agents are also important to study for their own sake. Indeed, as artificial agents are bound to participate more and more to the evolution of culture, it is crucial to better understand the dynamics of machine-generated cultural evolution. We here present a framework for simulating cultural evolution in populations of LLMs, allowing the manipulation of variables known to be important in cultural evolution, such as network structure, personality, and the way social information is aggregated and transformed. The software we developed for conducting these simulations is open-source and features an intuitive user-interface, which we hope will help to build bridges between the fields of cultural evolution and generative artificial intelligence.

\end{abstract}

\begin{multicols}{2}

\begin{figure*}
\begin{center}
\includegraphics[width=1\textwidth]{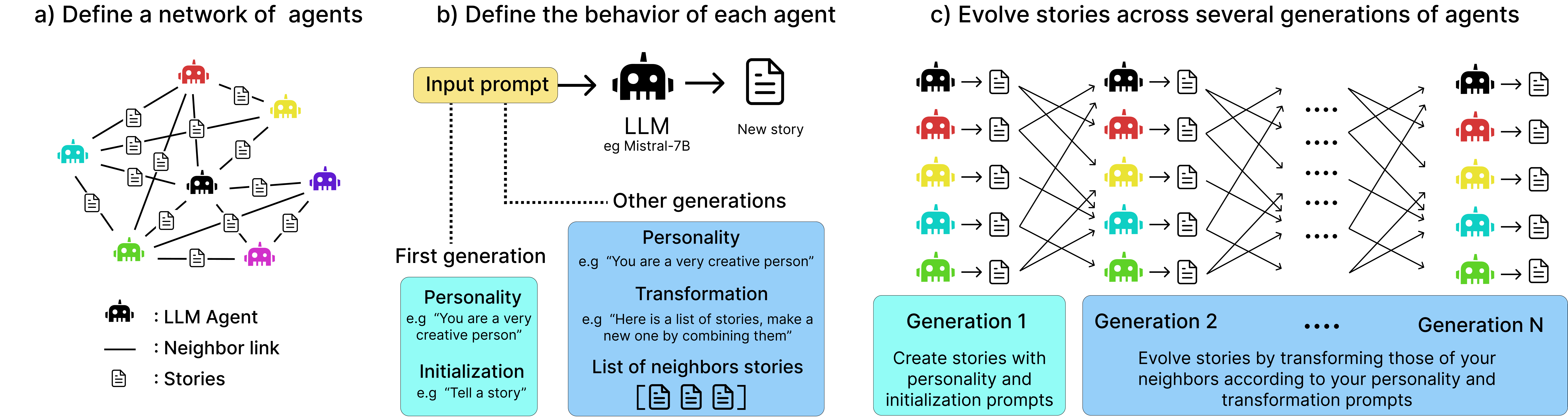}
\caption{\textbf{(a)} LLM agents are organized into networks wherein each agent interacts with neighboring agents by exchanging stories. \textbf{(b)} Each agent is assigned a specific personality and either initialization instructions (for the first generation) or transformation instructions (after the first generation), serving as prompts for generating new stories from their neighbors' narratives. \textbf{(c)} Once the network structure and agent characteristics are defined, we simulate the cultural evolution of texts across generations of agents. The simulation begins by prompting agents to initialize stories, after which we allow the narratives to evolve dynamically through interactions within the agent network} \label{intro_fig}
\end{center}
\end{figure*}

\section{Introduction}
Myths and stories are found in every human culture around the world. Research using phylogenetic methods has revealed that myths evolve in a way analogous to genes, progressively changing as some elements  are discarded and new elements added as they are passed through generations \cite{tehrani_phylogenetics_2017}. Some of these myths display a surprising stability, being retained with minor modifications though generations. For example, the Baku dream-eater demons, mythological creatures made of various animal parts, have been part of the Japanese folklore since at least the 15th century \cite{miton_cultural_2024}. Moreover, myths in distinct cultures seem to exhibit shared features, possibly pointing to convergent evolution. This is for example the case of historical myths (i.e., myths depicting events considered foundational for a given group), which have been found to possess many similar elements, such as narratives featuring the collective challenges faced in the history of the group \cite{sijilmassi_our_2024}. This raises questions regarding the factors that determine the cultural success, the stability, and the directions of evolution of myths and stories. 

More generally, providing causal explanations to the change of culture over time is the central aim of research in cultural evolution. Contributions to this question are generally seen as falling into two schools of thought: the first one, referred to as the Californian school, adopts the framework of gene-culture co-evolution \cite{boyd_culture_1988} \cite{cavalli-sforza_cultural_1981} \cite{richerson_not_2008} \cite{henrich_secret_2015}. This school emphasizes how evolved social learning mechanisms and transmission biases respectively contribute to the stability of culture and the direction of its evolution, and assume that the logic of natural selection applies to cultural traits as it does for genes \cite{richerson_not_2008}. The second school, referred to as the Parisian school, adopts the framework of Cultural Attraction Theory (CAT) \cite{Sperber1996-SPEECA}\cite{morin_how_2016} \cite{miton_cultural_2024}. Disagreeing on the central role attributed to selection, this view rather emphasizes how non-random transformations of cultural information during transmission events can explain how cultural traits progressively evolve toward stable forms, which are referred to as attractors. Taking inspiration from population genetics, the Californian school has successfully employed computational models to generate predictions, for instance about the effect of group size \cite{henrich_demography_2004}, network structure \cite{creanza_greater_2017} \cite{derex_divide_2018}, or transmission fidelity \cite{lewis_transmission_2012}. As for CAT, although computational models have also been proposed \cite{acerbi_culture_2021} \cite{mesoudi_cultural_2021} \cite{claidiere_how_2014}, they were mainly aimed at making conceptual points rather than to generate testable predictions, and attraction dynamics have thus been studied mainly through experiments \cite{kalish_iterated_2007} \cite{miton_motor_2020} \cite{miton_universal_2015} \cite{claidiere_cultural_2014}\cite{claidiere_cultural_2014} \cite{feher_novo_2009} and historical data analysis \cite{kelly_predictable_2024} \cite{dubourg_why_2022}.  

This prevalence of experimental and historical methods is likely attributable to the difficulty in capturing factors of attraction—such as evolved psychological mechanisms—that influence cultural evolution within agent-based or formal models.
Here, we propose that this limitation of agent-based models can be overcome by using generative artificial agents. In particular, using Large Language Models (LLMs) to simulate the evolution of linguistic culture appears fruitful, as we can expect those models to transform cultural information in realistic ways.

On top of allowing to generate hypotheses about the causes of cultural change in humans societies, studying the evolution of culture in populations of generative agents is also crucial to understand the dynamics of AI-generated cultural evolution. Indeed, although technology has long influenced cultural evolution, for example with the invention of the printing press, some authors have argued that the recent advances in generative algorithms are likely to result in an unprecedented shift in cultural evolution. The fact that algorithms are now participating in the creation of cultural traits sets us at the very beginning of an era of “machine culture”, defined as “culture mediated or generated by machines” \cite{brinkmann_machine_2023}. Therefore, studying the dynamics of machine-generated culture becomes highly important. Applying paradigms from Cultural Attraction theory to study LLMs has already began to generate precious insights about the bias exhibited by LLMs when transmitting linguistic content \cite{acerbi_large_2023}.

Overall, it appears that there is much to gain by using LLMs as agents to simulate the evolution of culture. So far, the fields of cultural evolution and generative artificial intelligence have remained quite disconnected, and facilitating exchanges between these fields seems promising for generating important findings. Some parallel works have investigated the use of Large Language Models as agents in multi-agent simulations, although these studies were not specifically designed to model cultural evolution \cite{park_generative_2023}\cite{vezhnevets_generative_2023}.

Here, we propose an open-source software for simulating the transmission and evolution of linguistic culture in populations of Large Language models. This software features an intuitive interface that allows to manipulate many variables of interest, such as the structure of the social network, the ways in which different sources of information are aggregated, or the personalities of the different agents. We also introduce several visualizations and measures that are useful for tracking the results of the simulations. We hope that this tool will pave the way for more interaction between the fields of cultural evolution and generative artificial intelligence. This software is open source and accessible at \href{https://github.com/jeremyperez2/LLM-Culture}{this link}.

\begin{figure*}%
  \centering%
  \includegraphics[width=1\linewidth]{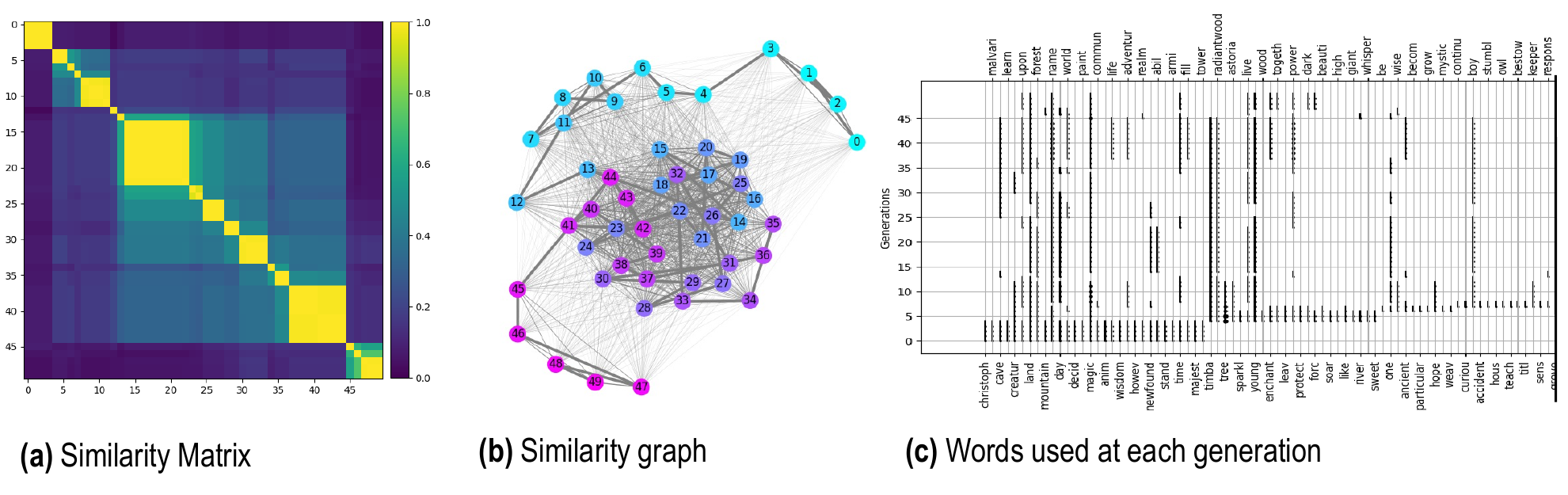}
\caption{Visualization of the evolution of the texts generated along a chain of 50 agents (i.e. 50 generations of one agent per generation). \textbf{(a)} The similarity matrix represents the semantic similarity between all stories generated. The color of the cell at row \textit{i}, column \textit{j} corresponds to the similarity of the stories \textit{i} and \textit{j}, which here corresponds to the stories generated at generation \textit{i} and generation \textit{j}. \textbf{(b)} The similarity graph is another way of visualizing the similarity between all generated stories. Each node corresponds to one story, and the distance between node is proportional to the semantic distance between corresponding stories. Stories generated at successive generations are linked by a wider edge and arer assigned similar colors. \textbf{(c)} We also visualize the words used at each generation. Words are along the x-axis, and generation along the y-axis. A dot at position \textit{(x,y)} means that the word \textit{x} was used in a story at generation \textit{y}. A line means that the corresponding word was used by two successive generations. We only display the first half of the figure here. The complete figure can be found in the Fig.~\ref{rotated_word_chains} of Appendix.}\label{Chain_vizualizations_fig}%
\end{figure*}%

\section{Methods}

\subsection{Description of the model}

The model presented here simulates the cultural evolution of linguistic content in a population of large language models (LLMs). Each agent can be seen as an independent instance of a LLM. The agent are arranged according to a specified social network structure (Fig.~\ref{intro_fig}.a). Network structures that can currently be generated with the software are detailed in the Appendix \ref{networks_details}. At the first generation, all agents are prompted with an Initialization Prompt. This prompt describes what kind of content the agents should generate. For example, one could chose “Tell me a story” as the initialization prompt. All agents then output an answer by passing the initialization prompt to their respective instance of the LLM. The agents then transmit stories to their neighbors (according to the specified network structure): each agent receives a new prompt, which is the concatenation of a Transformation Prompt and the list of stories produced by its neighbors at the previous generation (Fig.~\ref{intro_fig}.b). An example of a Transformation Prompt could be “Here are some stories. Make up a new one by combining two of them”. Agents can additionally be provided with a personality, which is always added at the top of their prompt. For example, a personality prompt could be “You are very imaginative”.  Agents may either all have the same personality or have different personalities. 

\subsection{Analysis}
\subsubsection{Similarity}
The main metric we use to analyze our results is the similarity between texts. To compute this metric, we first use TfidfVectorizers from scikit-learn \cite{pedregosa2011scikit} to convert texts into meaningful numerical representations. We then compute the cosine similarity \cite{han_2_2012} between all the texts generated, resulting in a similarity matrix of size $(N_{agents}$*$ N_{generations})$ X $(N_{agents}$*$ N_{generations})$, where the color of the cell (i,j) represents the similarity between story i and story j. 
 
From this similarity matrix, we extract more interpretable measures. The first one is the within-generation similarity, which captures how similar are the texts generated at a given generation with each other. The second one is the successive similarity, which represents the average similarity between the texts generated at a given generation and the texts generated at the previous generation. The last one is the similarity with the first generation, which is the average similarity between the texts generated at a given generation and the texts generated at the first generation. 


\subsubsection{Visualization}

We also provide two visualization techniques that offer qualitative insights into the generated data. 

\paragraph{Word chains}
We extract the key words from each text and represent their evolution though generations. To extract keywords from texts, we tokenize the text into words, remove common stopwords and non-alphanumeric tokens, calculate the frequency distribution of the remaining words, and select the top keywords based on their frequency. This allows to visualize which words are the most frequent, the most stable, or the most often reinvented.

\paragraph{Similarity network}

We also represent the similarity between generations using a graph network, where each node represents a generation of texts. The positioning of nodes is determined by a layout algorithm provided by the NetworkX library \cite{hagberg2008exploring}, arranging them based on their similarities and interconnections. Generations with the highest similarity are positioned closer together, and successive generations, represented by similar colors, are linked by thicker edges. This approach provides an intuitive depiction of the evolutionary dynamics of the generated content.

\section{Preliminary Results}

\begin{figure*}[t!]
\begin{center}
\includegraphics[width=1\textwidth]{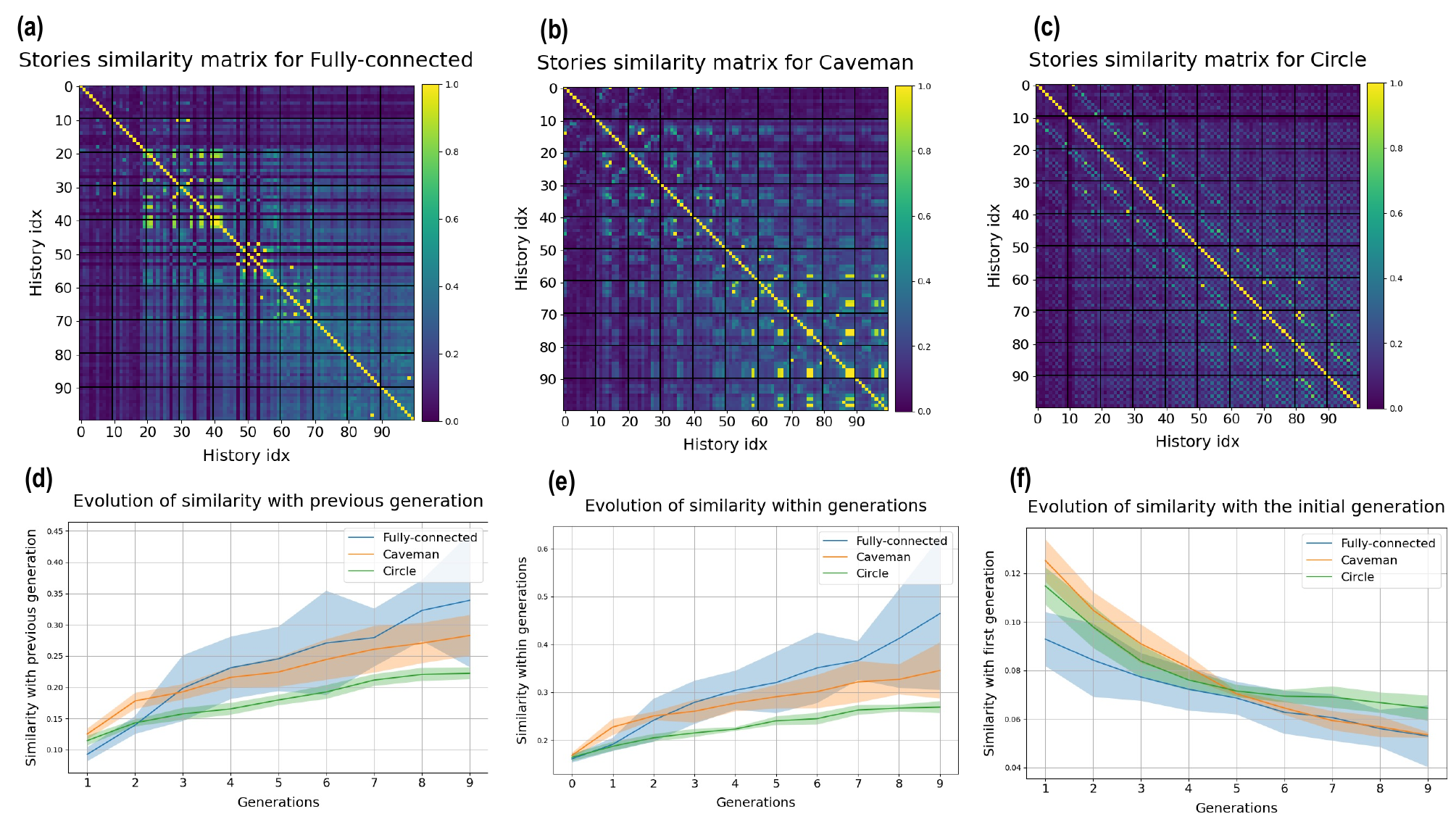}
\caption{\textbf{Effect of network structure :} Cultural dynamics of a population of 10 agents over 10 generations for three different types of network structures. \textbf{(a), (b), (c)} Similarity matrices for a fully-connected network, a caveman network with 2 cliques, and a circle network. The index of a story is defined as the agent's index (here between 0 and 10) + $N_{agents}$ * generation-index. For example, the story 3 belongs to the first generation, and story 13 to the second generation. The color of the cell at row \textit{i} and column \textit{j} represents the semantic similarity between stories \textit{i} and \textit{j}. Black lines mark the separation between generations. \textbf{(d)} Evolution of the average similarity between stories produced at a generation and stories at the generation just before.\textbf{(e)} Evolution of the average similarity between each pair of stories produced at a given generation.\textbf{(f)} Evolution of the average similarity between stories produced a given generation and stories produced at the first generation. Lines represent averages over 5 simulations, and the filled areas represent the standard deviations.} \label{structure}
\end{center}
\end{figure*}

The results presented in this paper are preliminary, and are mainly meant to illustrate how one can manipulate with the different variables to see their effects on the measures described above. However, they do already offer some interesting insights.

\subsection{Transmission chain}\label{transmission_chain}

We first illustrate the dynamics of the model using a linear transmission chain of 50 agents (i.e one agent per generation for 50 generations). The initialization and transmission prompts used are provided in the Section~\ref{prompts_content} of Appendix. The agents were not assigned any personality for this experiment (personality prompt is empty). 
Fig.~\ref{Chain_vizualizations_fig}.a shows the similarity matrix for this chain. We can notice that stories seem to evolve in a punctuated manner: there is an alternation of phases where stories are transmitted with no modifications, and phases where stories are modified. 
These dynamics recall previous observations from experiments and modelling work in cultural evolution \cite{kolodny_evolution_2015} \cite{valverde_punctuated_2015} \cite{mokyr_punctuated_1990} \cite{miu_innovation_2018}, where cultural information was found to evolve in a sequence of bursts and stasis.
The transition matrix also suggests the existence of hierarchically structured clusters of stories that are not identical but share some degree of similarity. 
Fig.~\ref{Chain_vizualizations_fig}.b shows the graph of similarity of this chain, which allows to visualize how stories progressively evolve through a semantic space. Looking at this graph suggests that stories got "trapped" in a specific part of the semantic space for a while (from generation 14 to 41), possibly indicating the presence of an attractor. 
Fig.~\ref{Chain_vizualizations_fig}.c shows the chain-of-words representation for this chain. This visualization allows to notice that some words are very stable, being kept almost throughout the chain, such as the word "magic", while some words only last for the first few generations before being lost, such as the word "learn". 

\begin{figure*}[t!]
\begin{center}
\includegraphics[width=1\textwidth]{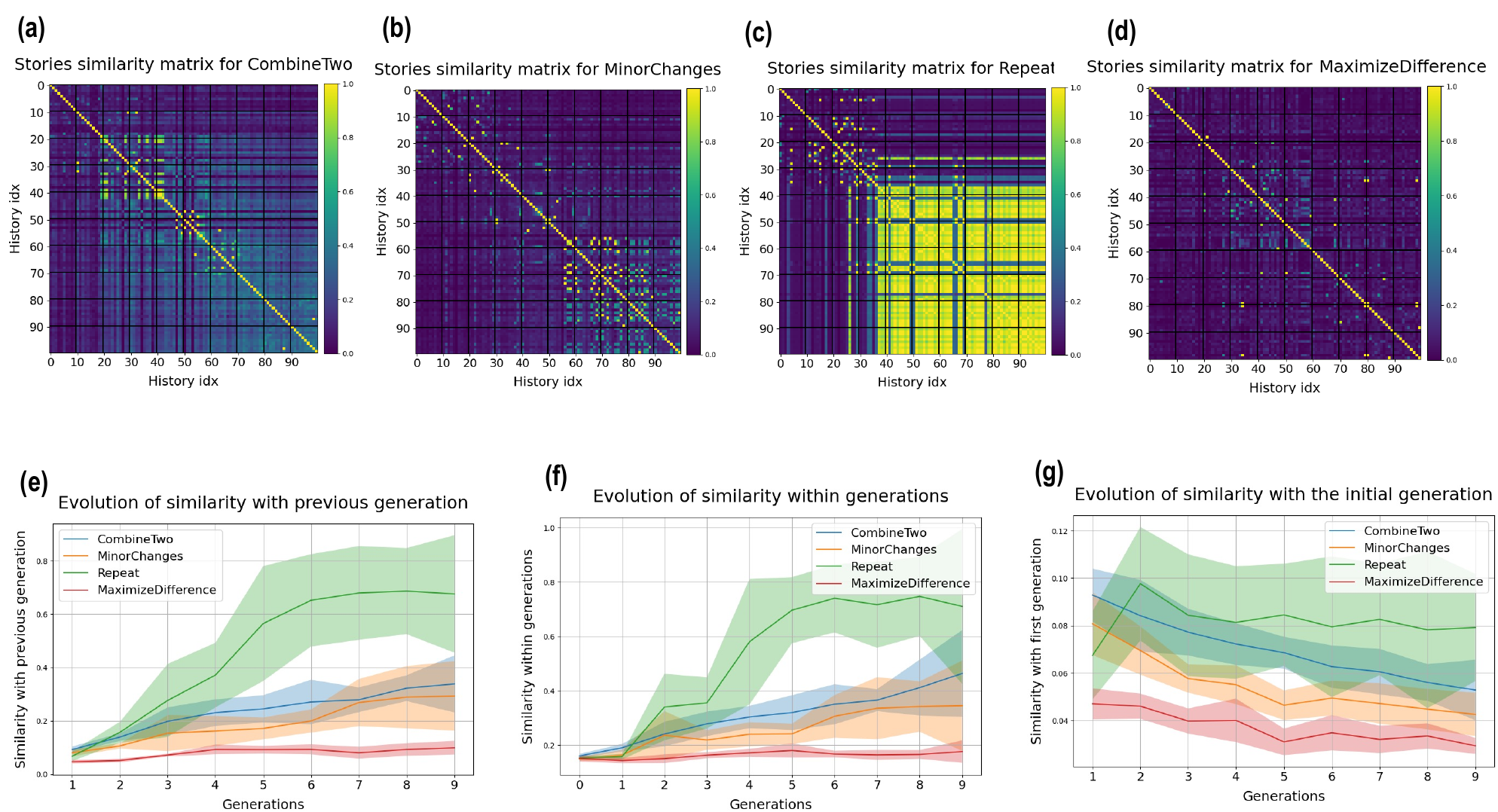}
\caption{\textbf{Effect of transmission prompts :} Cultural dynamics of a population of 10 agents over 10 generations for four different transmission prompts. \textbf{(a), (b), (c), (d)} Similarity matrices for the "CombineTwo", "MinorChanges", "Repeat" and "MaximizeDifference" transformation prompts. \textbf{(e)} Evolution of the average similarity between stories produced at a generation and stories at the generation just before.\textbf{(f)} Evolution of the average similarity between each pair of stories produced at a given generation.\textbf{(g)} Evolution of the average similarity between stories produced a given generation and stories produced at the first generation. Lines represent averages over 5 simulations, and the filled areas represent the standard deviations.} \label{prompts}
\end{center}
\end{figure*}

\subsection{Effect of network structure}\label{networks_section}

We first manipulated the network structure by running 5 simulations of 10 agents for 10 generations, for each network structure (Fig~.\ref{structure}). We compared three network structures: a fully-connected network, a circle network,  and a caveman network with 2 cliques (see Fig~.S\ref{networks_fig} of the Appendix). We only show one of the five similarity matrices for each network structure, and provide all matrices in Appendix \ref{additional_results}.  Qualitatively, the similarity matrices reveals distinct patterns for each network structure: while similarity appears to increase homogeneously for the fully-connected network, it does so in a clustered manner for the caveman and circle networks. Looking at the evolution of the similarity measures, we observe that for all structures, stories get more and more similar to stories of the same generation and of the previous generation, indicating a progressive homogenisation of the population. We also observe that stories get more and more dissimlar with stories from the first generation. For all measures, these dynamics seem to happen at a faster rate for the fully-connected structure than for the caveman network, itself faster than the circle network. These results replicate findings from previous work in network science and cultural evolution, where models and experiments reveal that more efficient networks (that is, networks with smaller average path length) lead to quicker diffusion of information and thus lower diversity  \cite{derex_partial_2016} \cite{derex_divide_2018} \cite{nisioti_social_2022} \cite{lazer_network_2007}.

\subsection{Effect of transformation prompt}

We then manipulated the transformation prompt by running 5 simulations of 10 agents for 10 iterations (Fig.~\ref{prompts}). We compared the effect of four different transformation prompts referred to as "Combine2", "MinorChanges", "Repeat" and "MaximizeDifference". The exact content of these prompts is provided in Appendix ~\ref{prompts_content}. 
These dynamics reveal that clear differences can be observed, and those differences are aligned with what would be expected. For example, the Repeat prompt leads to high within- and between- generation similarity, while these metrics are low for the MaximizeDifference prompt. Apart for the MaximizeDifference condition, all conditions show a gradual increase in the simlilarity with previous generation and within-generation, indicating a progressive homogenisation of the cultural content. Lastly, for all conditions but the Repeat condition, similarity with the initial generation appears to decrease over time. This is in particular interesting for the MaximizeDifference condition, as the second generation was already instructed to maximize the difference with the first generation.

\subsection{Effect of different personalities}

\begin{figure*}[t!]
\begin{center}
\includegraphics[width=1\textwidth]{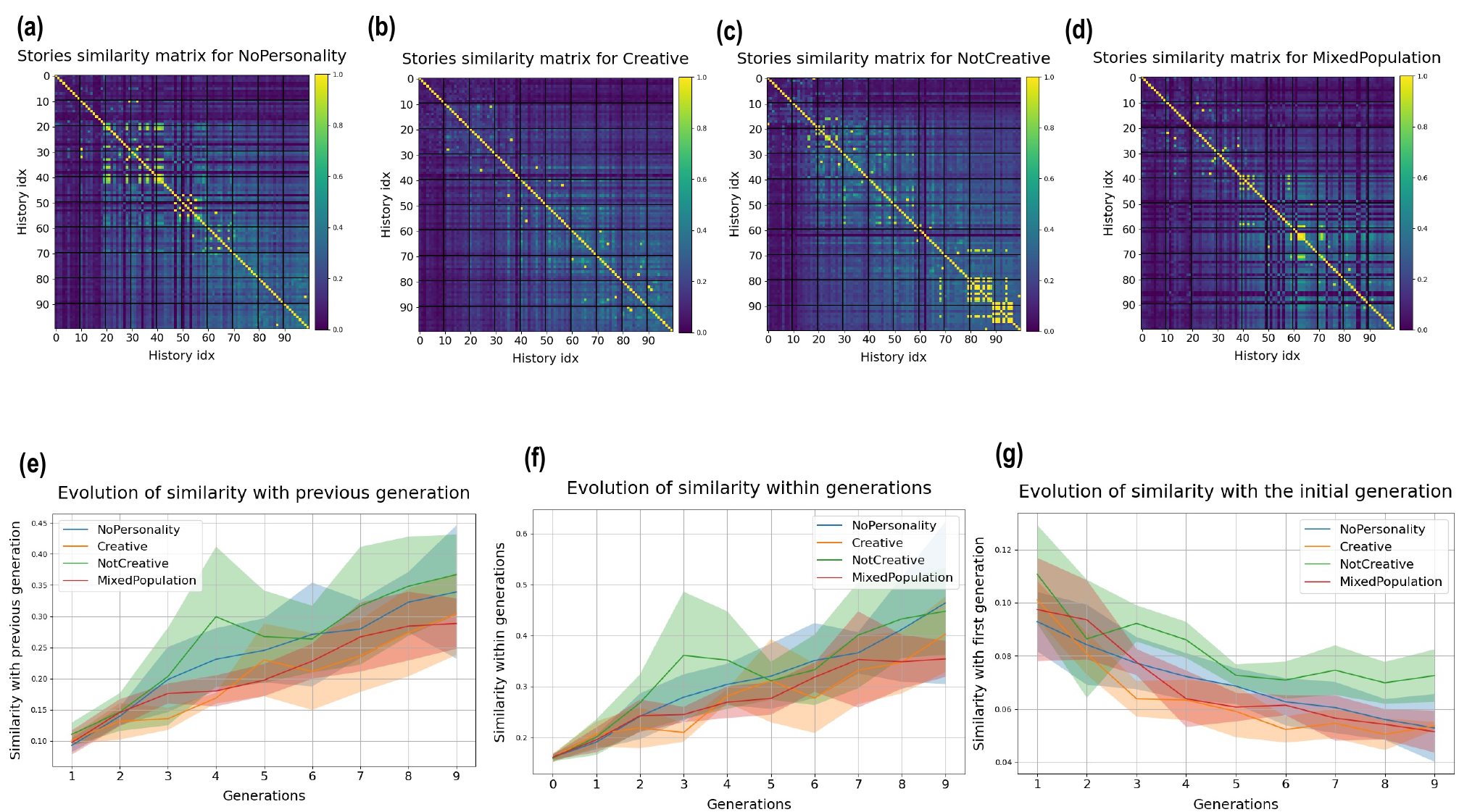}
\caption{\textbf{Effect of personalities :} Cultural dynamics of a population of 10 agents over 10 generations for four different personnalities. \textbf{(a), (b), (c), (d)} Similarity matrices for  "NoPersonality", "Creative", "NotCreative" and Mixed population of "Creative" and "NotCreative". \textbf{(e)} Evolution of the average similarity between stories produced at a generation and stories at the generation just before.\textbf{(f)} Evolution of the average similarity between each pair of stories produced at a given generation.\textbf{(g)} Evolution of the average similarity between stories produced a given generation and stories produced at the first generation. Lines represent averages over 5 simulations, and the filled areas represent the standard deviations} \label{personalities}
\end{center}
\end{figure*}

We then manipulated the personality assigned to agents by running 5 simulations of 10 agents for 10 generations (Fig.~\ref{personalities}).  We compared 4 different conditions. In the first one, agents were not assigned any personality. In the second one, all agents were assigned the personality "Creative". In the third one, all agents were assigned the personality "NotCreative". In the fourth one, half of the agents were assigned the personality "Creative" and the other half was assigned the personality "NotCreative". Details of the personality prompts can be found in Appendix \ref{prompts_content}. Although performing more simulations would be necessary to draw conclusions, it does seem that different personalities differentially impact cultural dynamics. In particular, our results suggest that agents in the Creative condition generate more variation than agents in the NotCreative condition: indeed, the NotCreative appears to exhibit higher similarity with the first generation, between successive generations, and within generations compared to the Creative condition. Also interesting is the fact the the MixedPopulation condtion seems to be more similar to the Creative condition than to the NotCreative condition, even though it contains Creative and NotCreative agents in equal proportions. This suggests that the cultural dynamics of an heterogeneous population may not reflect the proportions of the different personalities it contains.

\section{Discussion}

We introduced a framework for modelling cultural evolution in population of LLMs agents, along with several metrics and visualizations that allows to extract qualitative and quantitative insights about the resulting cultural dynamics. We presented how this framework allows to manipulate several variables known to impact cultural evolution, namely the network structure, the way of transforming previous social information, and agents’ personalities. Although the results presented in this paper are still preliminary, they already reveal several insights. First, the fact our simulations replicate several findings from empirical and  theoretical work in cultural evolution confirms that LLMs-based multi-agent models are an adequate tool for generating hypotheses about the dynamics of human culture. It also suggests that results from studies of human-generated culture may apply to machine-generated culture. Moreover, we found that manipulating personality and social information transformation appears to have significant impact on the observed dynamics. This suggests that using LLMs to study cultural evolution is promising, as these variables have so far been difficult to model using traditional modelling tools. It also indicates that the dynamics of machine-generated culture will likely be influenced by the specific way in which generative agents are instructed to use pre-existing social information, and which personality traits to emulate. 

These results were preliminary and should mainly be seen as proof of concept. Future work will involve performing more systematic and rigorous analyses of the effect of the different variables on cultural evolution. It would also be interesting to compare how different groups, all starting with the same story, evolve over time. 

Although Large Language models are a useful proxy of human behavior, some precautions must be taken when generalizing results of such simulation to human behavior. First, because of the biases in their training set, LLMs are mainly representative of western culture \cite{atari_xue_park_blasi_henrich_2023}. As such, they may not be able to capture the specific dynamics of other cultures. A potential way to alleviate that is to use models that were trained on more representative datasets, such as the BLOOM model whose training data includes 46 different languages \cite{bigscience_workshop_bloom_2023}. More generally, precisely evaluating to what extent current models miss some aspects of non-western cultures is a major direction for future work on machine-generated cultural evolution. A second limitation to keep in mind is that in some situations, LLMs may struggle to role-play and impersonate specific characters \cite{kovač2024stick}. Better understanding to what extent they take into account their assigned personality when creating cultural content is a crucial step in order to derive hypotheses about human culture using such models. 

Another important difference between this model and human cultural evolution is that the latter is grounded in a physical environment. As such, the success of cultural traits not only depend on the cognitive mechanisms of the agents transmitting it, but also on their relationship with the environment of the agents \cite{thompson_human_2021}. For example, beliefs in moralizing gods have been found to be predicted with high accuracy by the ecological and historical context \cite{botero2014ecology}. Extending this model to include interaction with a physical environment therefore seems like a very promising direction. 

Finally, the current model assumes that personalities are fixed and stable through time. However in humans, there are feedback loops between the cultural traits expressed in a group and the personalities, values and preferences of its member \cite{smaldino_niche_2019} \cite{higgins_culture_2008}. Including such interactions would be fruitful to model phenomena such as opinion dynamics and polarization. 

More generally, although we here focused on the dynamics of collective creation, this framework is suited to explore other questions related to collective behavior, including opinion dynamics, collective innovation and the evolution of language.

Overall, we have sketched how using generative agents to simulate cultural evolution may be promising, both for generating hypotheses about human cultural evolution and for better understanding the dynamics of machine-generated culture. We hope that making this tool easy to use (see usage details in Appendix \ref{user_interface}) and accessible to the scientific community will foster more exchanges between the fields of cultural evolution and generative artificial intelligence.

\section*{Acknowledgements}

This research was partially funded by the French National Research Agency (\href{https://anr.fr/}{ANR}, project ECOCURL, Grant ANR-20-CE23-0006).
This work benefitted from access to the Jean Zay (Idris) supercomputer associated with the Genci grant A0151011996.
This research originated as a project from the Hackathon \href{https://sites.google.com/view/hack1robo/accueil}{Hack1Robo}.
We also thank Maxime Derex for participating in helpful discussions.

\end{multicols}

\bibliographystyle{plain}
\bibliography{RefGlobalAICulture.bib}

\pagebreak

\begin{appendices}

\renewcommand{\figurename}{Supplementary Figure}
\setcounter{figure}{0}

\section{Simulation details}

\paragraph{LLM used}
The results presented in this paper were generated using MistralOrca \cite{lian2023mistralorca1}, which is a version of openAI's Mistral-7B Model, instruct-tuned on Filtered OpenOrcaV1 GPT-4 Dataset. We used a quantized GGUFv2 version, quantized using the method "GGML-TYPE-Q4-K". 

\paragraph{Prompts used for the experiments}\label{prompts_content}

For the chain of 50 agents (Section~\ref{transmission_chain}), the initialization prompt used was "Imagine that you are telling a story to your kid. What would that story be? Just output the story, nothing else." The transformation prompt used was "Here is one or more stories you were told as a kid. It is now your turn to tell a story at your kid. Tell that story. Write only one story. Do not output anything else.". 

For all other simulations, the initialization prompt was "Tell me a story". Except when manipulating the transformation prompt, the transmission prompt for all simulations was CombineTwo. Here is the complete list of transformation prompts used in our experiments :

\begin{itemize}
    \item \textbf{CombineTwo (default)}: 
    \begin{quote}
        "You will receive stories. Pick the two stories you prefer, and create a story that is combination of these two stories. Just output your story, don’t write anything else."
    \end{quote}
    
    \item \textbf{MinorChanges}: 
    \begin{quote}
        "You will receive a list of one or more stories. Create a new story by making some minor changes to one of those stories. Just output one story, do not output anything else."
    \end{quote}
    
    \item \textbf{Repeat}: 
    \begin{quote}
        "You will receive stories. Select only one of these stories, and repeat it. Just output the story, don’t write anything else."
    \end{quote}
    
    \item \textbf{MaximizeDifference}: 
    \begin{quote}
        "You will receive stories. Create a story that is as different as possible from the stories you received. Just output your story, nothing else."
    \end{quote}
\end{itemize}

Except when manipulating the personality, the personality prompt was an empty string for all simulations. When manipulating the personalities, the prompts were:

\begin{itemize}
    \item \textbf{Creative}: 
    \begin{quote}
        "For what follows, pretend that you are a very creative person.”
    \end{quote}

    \item \textbf{NotCreative}: 
    \begin{quote}
        “For what follows, pretend that you are not a very creative person.”
    \end{quote}
    
\end{itemize}
\paragraph{Network structure}\label{networks_details}

\begin{figure*}[t!]
\begin{center}
\includegraphics[width=1\textwidth]{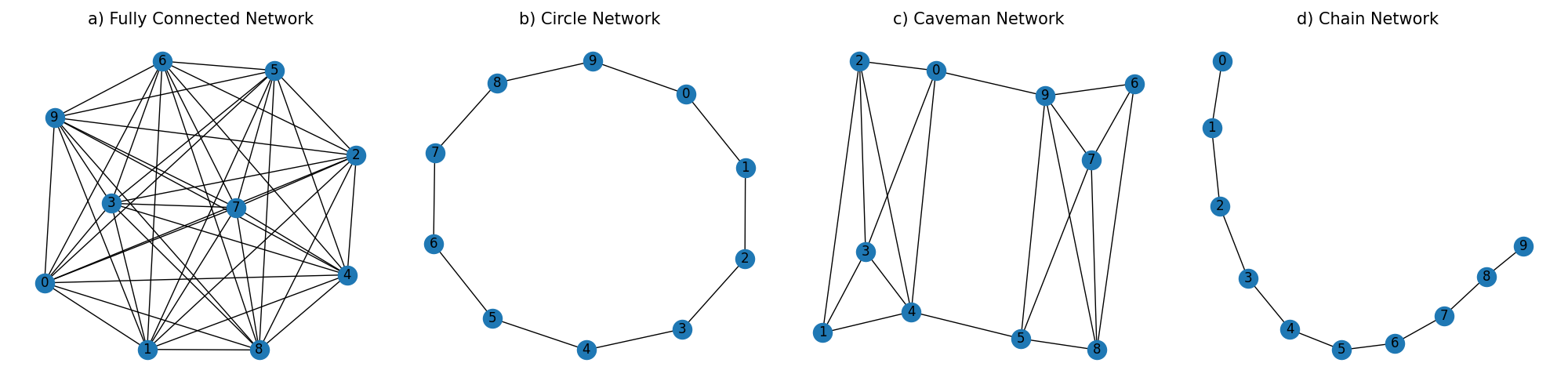}
\caption{Different types of networks used in our experiments with 10 agents. In a fully connected network \textbf{(a)}, every agent is directly connected to every other agent, enabling efficient dissemination of information. In a circular network \textbf{(b)}, agents are connected in a circular fashion, thereby forming a closed loop where stories flow sequentially around the circle. The caveman network \textbf{(c)} consists of agents organized into cliques, with our example showcasing two cliques of 5 agents each. In the chain network \textbf{(d)}, agents are arranged linearly, and content generation happens sequentially, meaning that agents only generate a story after receiving the story of the previous agent in the chain.}\label{networks_fig}
\end{center}
\end{figure*}

Unless the network structure was the manipulated variable, the default structure was a fully-connected network (Supplementary Fig~.\ref{networks_fig}.a). We also used a chain network (Supplementary Fig~.\ref{networks_fig}.d) in Section~\ref{transmission_chain}, as well as circle (Supplementary Fig~.\ref{networks_fig}.b) and caveman (Supplementary Fig~.\ref{networks_fig}.d) networks in Section~\ref{networks_section}.

\pagebreak

\section{User Interface}\label{user_interface}

\begin{figure*}
\begin{center}
\includegraphics[width=1\textwidth]{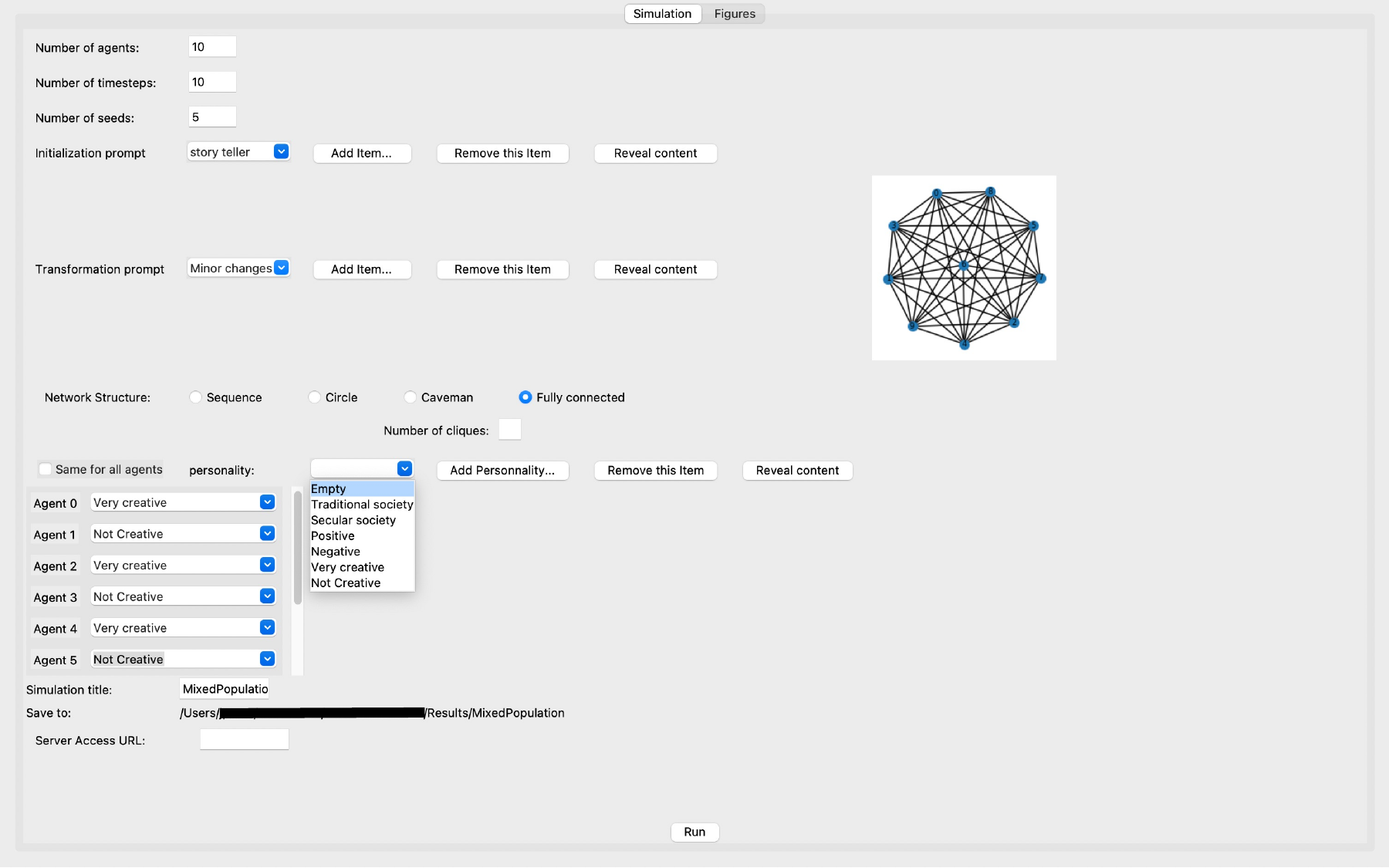}
\caption{Screenshot of the Graphical User Interface} \label{figS1}
\end{center}
\end{figure*}

To allow researchers to easily use our model, we developed an intuitive user interface that allows to manipulate the variables of interest and generate figures (see Supplementary Figure \ref{figS1} ) . In the Simulation panel, users can set these variables. Starting from the top of the panel, users can select how many agents they wish to simulate (“Number of agents”), how many generations to simulate (“Number of generations”), and how many times to repeat the simulations ("Number of seeds"). Users can also select which initialization prompt and transformation prompt to use (“Initialization prompt” and “Transformation prompt”). Users may add a new prompts by clicking on the “Add prompt…” button. This opens a new window in which they can set the name and content of the prompt. Users can also select which network structure to use for the simulation. When “sequence” is selected, agents are arranged in a chain, and each agents waits to receive the output of the previous agent in the chain before generating its own output. The other options are "Circle", which creates a cyclic graph, "Caveman", which creates a connected-caveman network \cite{watts_networks_1999} (in which case the user must also select the number of cliques), and a "Fully-Connected", which creates a fully-connected network. The structure of the chosen graph is displayed on the right side of the window. Finally, users can also select the personalities to assign to the agents. A checkbox allows to indicate whether all agents should have the same personality. If it is the case, users can select the corresponding personality in the drop-down menu. Like for the prompts, users can add new personalities to the list. If agents have different personalities, the users can assign each agent its personality in the corresponding list. 
On top of these parameters, users should give a name to the simulation, which will be the name of the folder in which results are saved. Users should also specify the URL to send the request to in order to get answers from the LLM. We generated such an URL using a library from oogabooga (https://github.com/oobabooga/text-generation-webui), and provide details of this procedure in the README file of \href{https://github.com/jeremyperez2/LLM-Culture}{our github repository}. 
Once all of these parameters have been set, users can run the simulations by clicking on "Run". Once the simulation has terminated, figures will be generated and displayed in the Figures tab of the GUI, as well as being stored in the result folder. 

\pagebreak

\section{Additional metrics and results}

\subsection{Additional metrics }

\paragraph{Creativity}
To obtain the creativity of a text, we compute the average semantic distance between all pairs of words in this text. This way of measuring creativity is supported by recent studies \cite{kenett_what_2019} \cite{johnson_divergent_2023}. 

\paragraph{Positivity and subjectivity} 

We used TextBlob \cite{loria2018textblob} to perform sentiment analysis on the content generated. In particular, we used their measures of positivity and subjectivity and looked at their evolution through generations.

\subsection{Additional results} \label{additional_results}

The evolution of those additional metrics though generations is shown in Supplementary Figure \ref{supplementary_structures}, Supplementary Figure \ref{supplementary_prompts} and Supplementary Figure \ref{supplementary_personalities}

\begin{figure*}[t!]
           
            \centering
            \includegraphics[width=\linewidth]{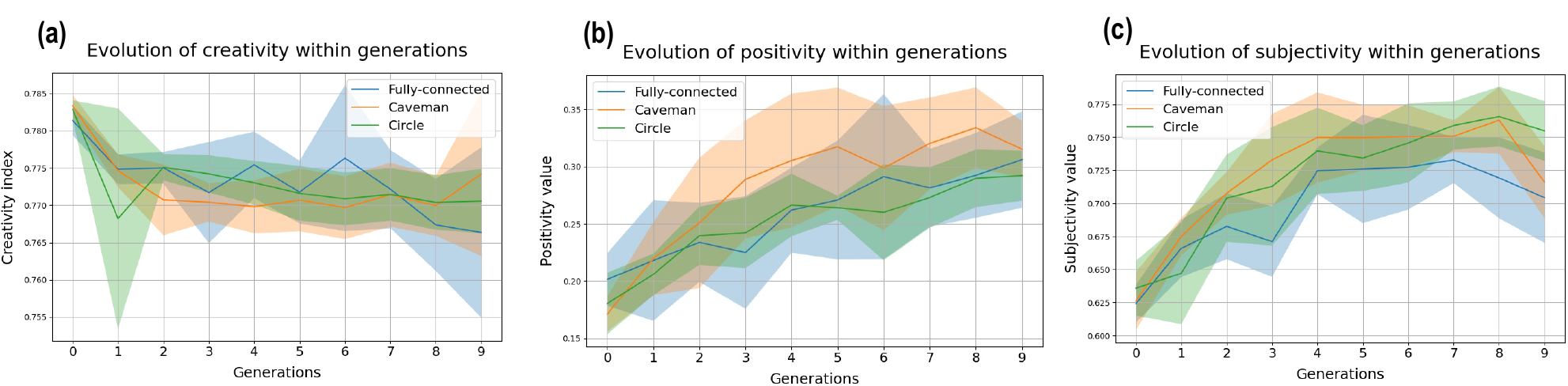}
           
    \caption{Evolution of additional metrics for four different network structures}
    \label{supplementary_structures}
\end{figure*}

\begin{figure*}[t!]
           
            \centering
            \includegraphics[width=\linewidth]{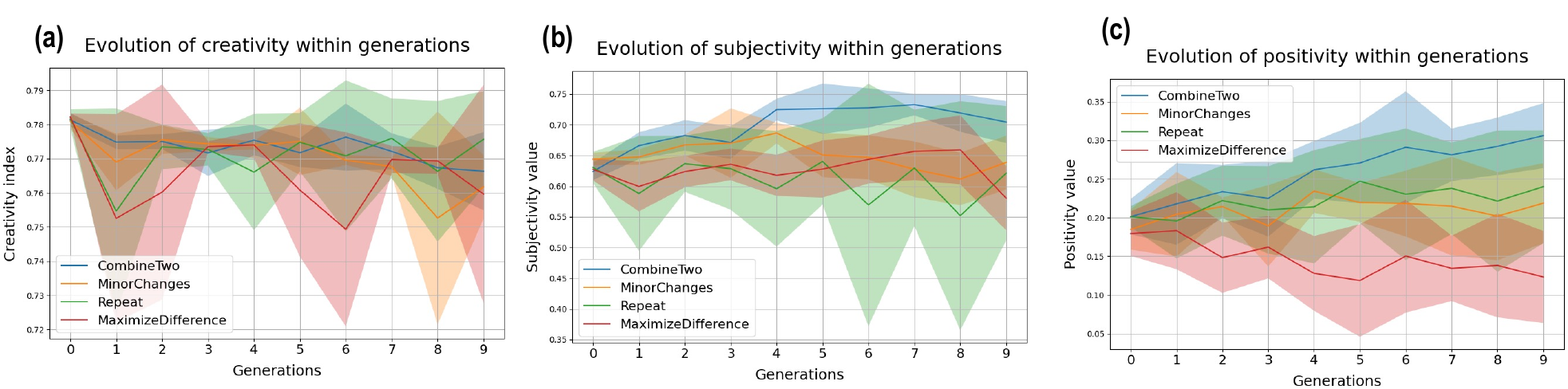}
           
    \caption{Evolution of additional metrics for four different transformation prompts: Combine2, MinorChanges, Repeat, and MaximizeDifference}
    \label{supplementary_prompts}
\end{figure*}

\begin{figure*}[t!]
           
            \centering
            \includegraphics[width=\linewidth]{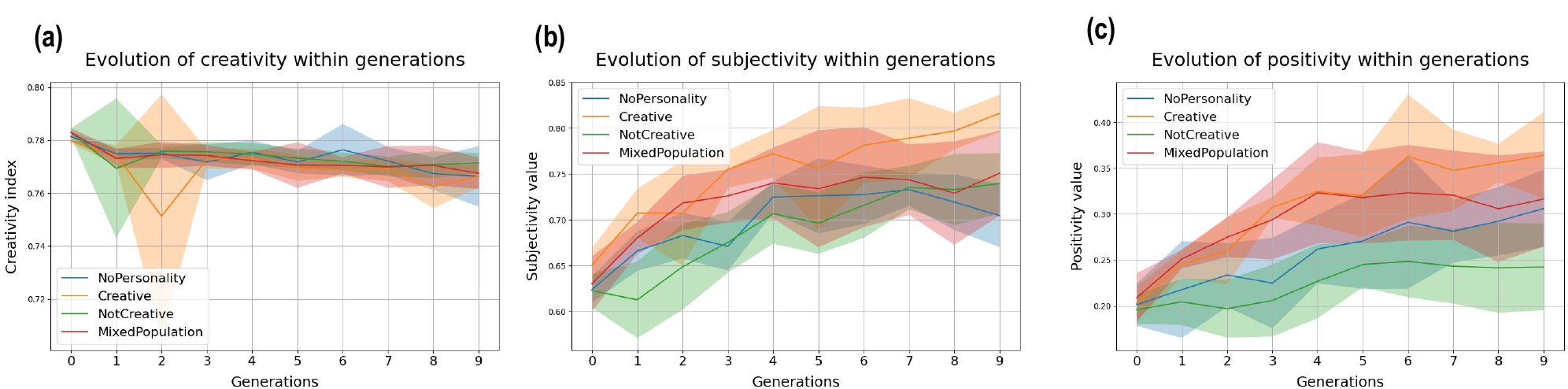}
           
    \caption{Evolution of additional metrics for different personalities:  NoPersonality, Creative, NotCreative, Mixed population of Creative and NotCreative}
    \label{supplementary_personalities}
\end{figure*}

\begin{figure*}[t!]
           
            \centering
            \includegraphics[width=\linewidth]{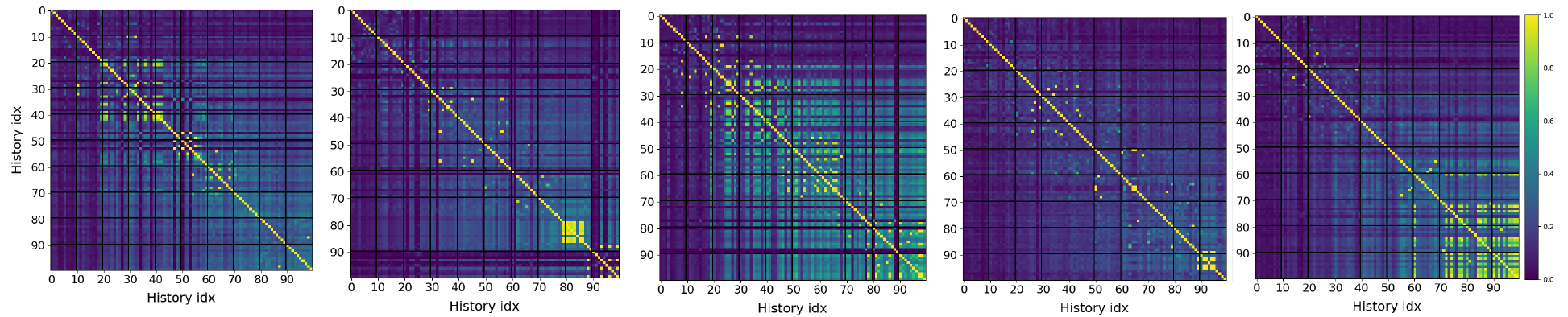}
           
    \caption{All similarity Matrices for the Fully-Connected Network.}
    \label{all_matrices_fc}
\end{figure*}

\begin{figure*}[t!]
           
            \centering
            \includegraphics[width=\linewidth]{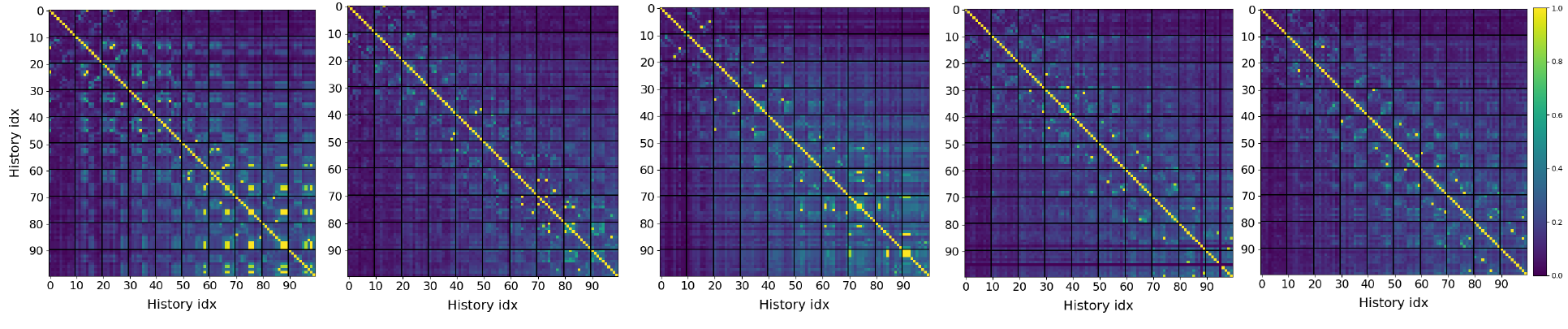}
           
    \caption{All similarity Matrices for the Caveman Network}
    \label{all_matrices_caveman}
\end{figure*}

\begin{figure*}[t!]
           
            \centering
            \includegraphics[width=\linewidth]{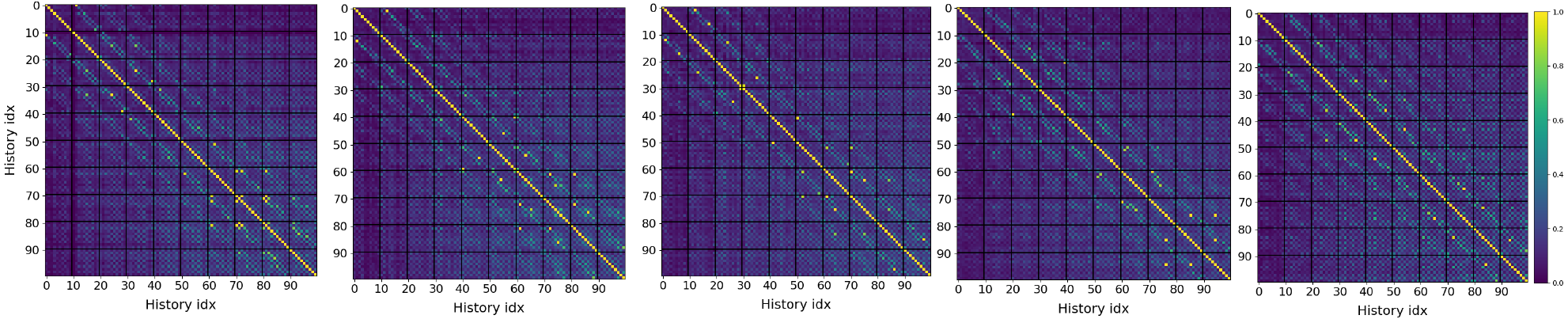}
           
    \caption{All similarity Matrices for the Circle Network}
    \label{all_matrices_circle}
\end{figure*}

\begin{figure*}[t!]
           
            \centering
            \includegraphics[width=\linewidth]{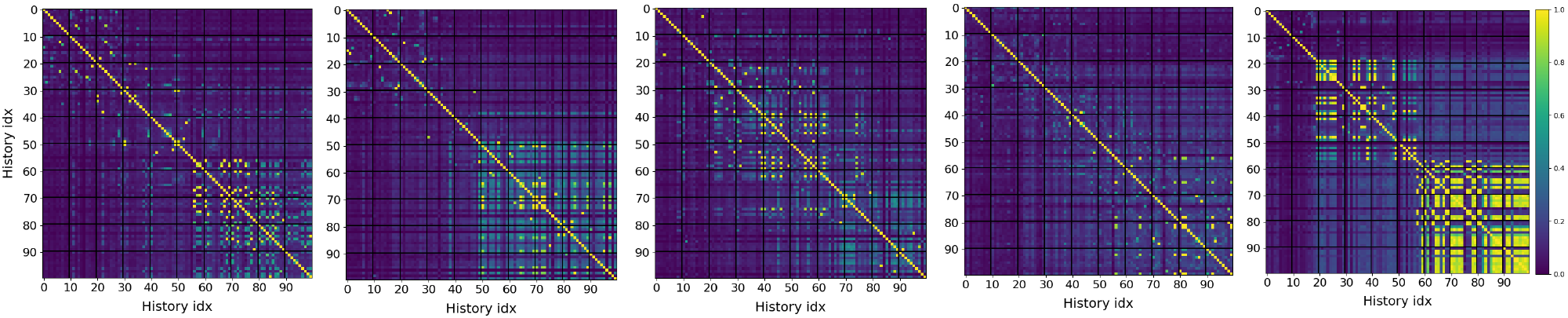}
           
    \caption{All similarity Matrices for the MinorChanges transformation prompt}
    \label{all_matrices_minor}
\end{figure*}

\begin{figure*}[t!]
           
            \centering
            \includegraphics[width=\linewidth]{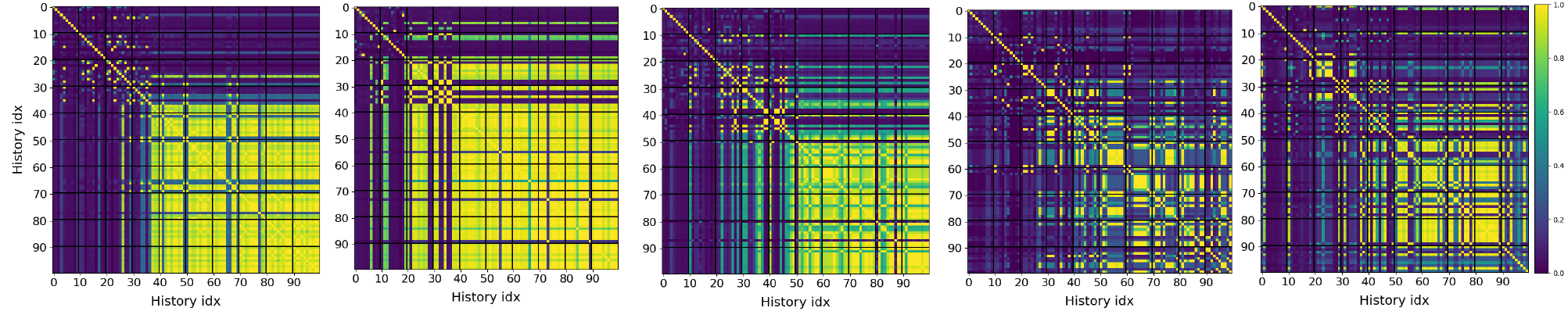}
           
    \caption{All similarity Matrices for the Repeat Prompt}
    \label{all_matrices_repeat}
\end{figure*}

\begin{figure*}[t!]
           
            \centering
            \includegraphics[width=\linewidth]{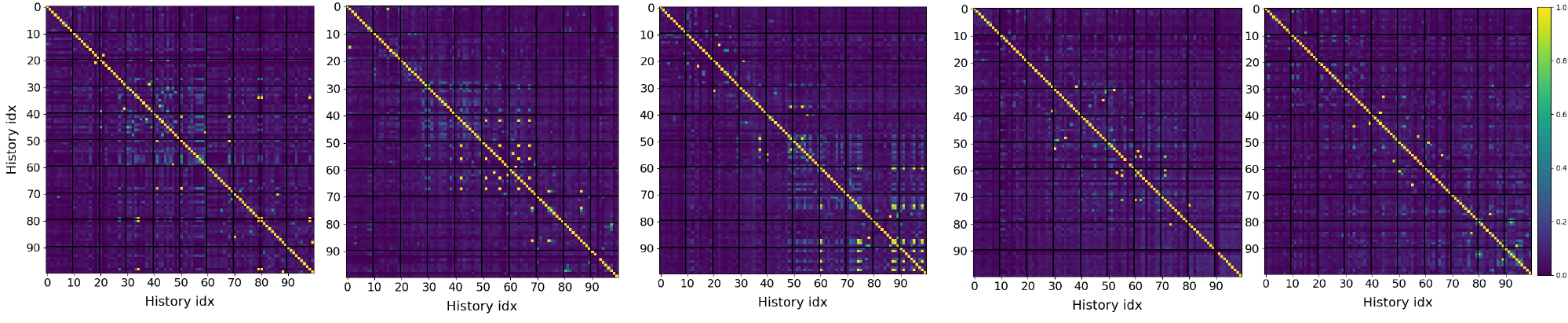}
           
    \caption{All similarity Matrices for the MaximizeDifference transformation Prompt}
    \label{all_matrices_difference}
\end{figure*}   

\begin{figure*}[t!]
           
            \centering
            \includegraphics[width=\linewidth]{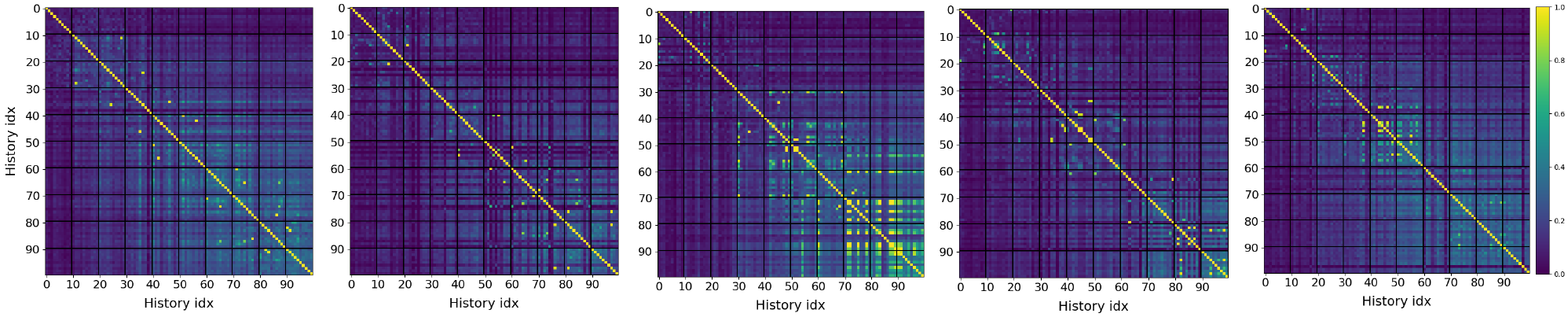}
           
    \caption{All similarity Matrices for the Creative personality}
    \label{all_matrices_creative}
\end{figure*} 

\begin{figure*}[t!]
           
            \centering
            \includegraphics[width=\linewidth]{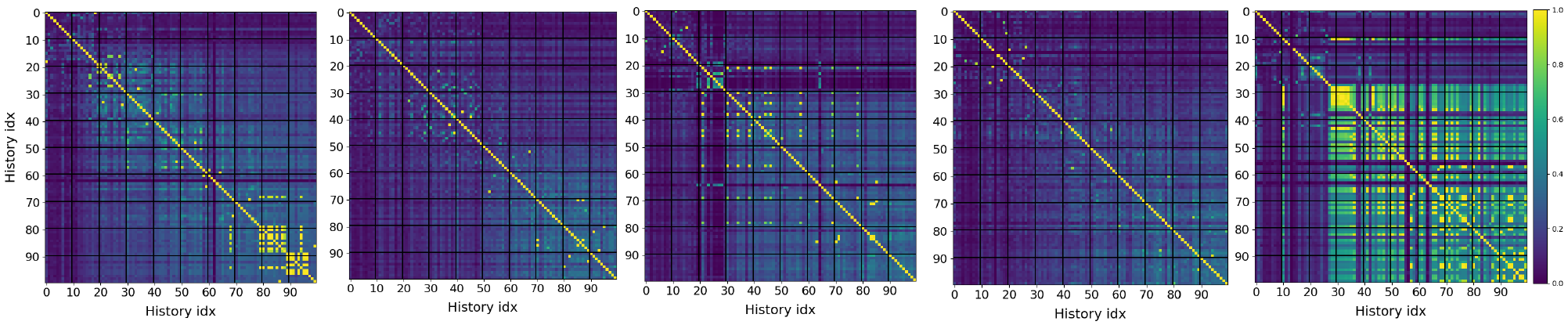}
           
    \caption{All similarity Matrices for the NotCreative personality}
    \label{all_matrices_notCreative}
\end{figure*} 

\begin{figure*}[t!]
           
            \centering
            \includegraphics[width=\linewidth]{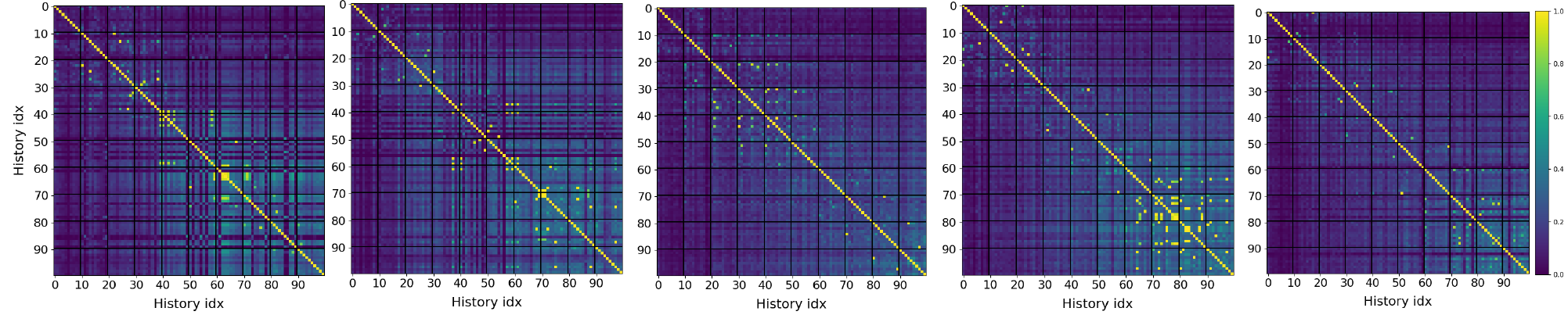}
           
    \caption{All similarity Matrices for the Mixed Population}
    \label{all_matrices_mixedPop}
\end{figure*} 

\begin{figure*}
\begin{center}
\includegraphics[width=0.45\textwidth]{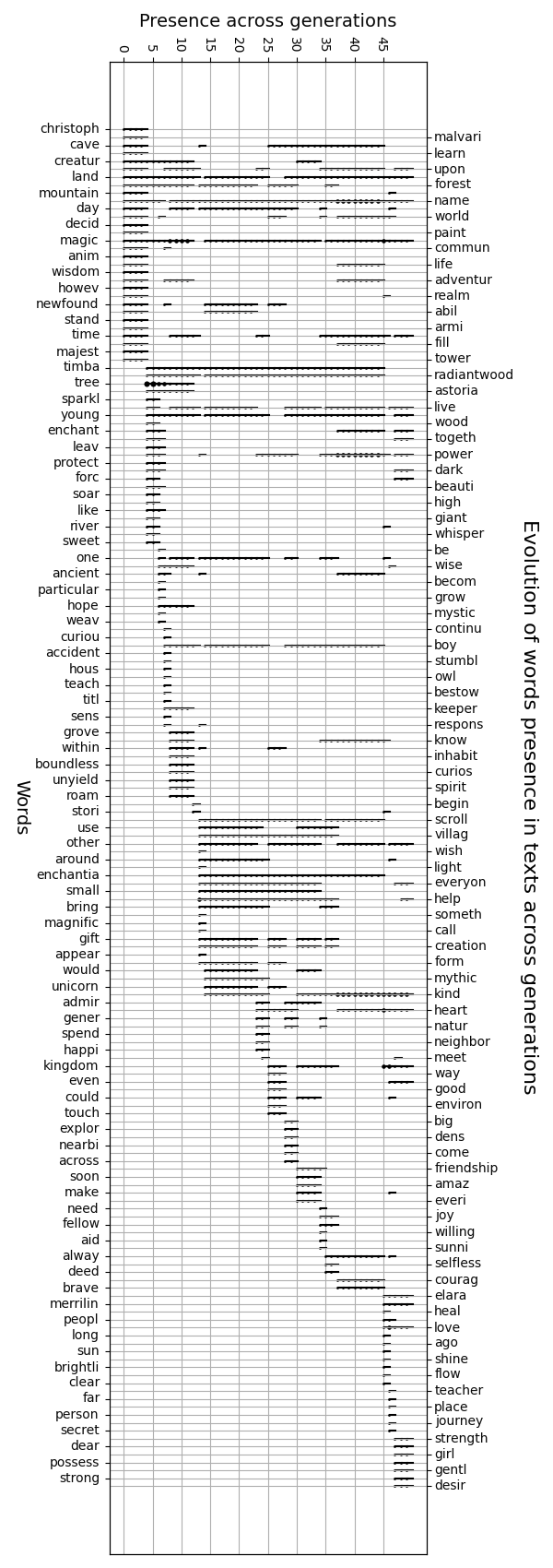}
\caption{Complete rotated word chains plot of fig~.\ref{Chain_vizualizations_fig}} \label{rotated_word_chains}
\end{center}
\end{figure*}

\end{appendices}

\end{document}